\documentstyle[11pt,aaspp]{article}
\input{psfig}
\eqsecnum
\def\beq{\begin{equation}}
\def\eeq{\end{equation}}
\def\ref{\reference}
\begin{document}
\title{Torque Reversal in Accretion-Powered X-ray Pulsars}
\author{Insu Yi$^1$, J. Craig Wheeler$^2$, and Ethan T. Vishniac$^2$}
\affil{$^1$Institute for Advanced Study, Princeton, NJ 08540; yi@sns.ias.edu}
\affil{$^2$Astronomy Department, University of Texas, Austin, Texas 78712;
ethan@grendel.as.utexas.edu, wheel@astro.as.utexas.edu} 

\begin{abstract}

Accretion-powered X-ray pulsars 4U 1626-67, GX 1+4, and OAO 1657-415 have 
recently shown puzzling torque reversals. These reversals are characterized 
by short time scales, on the order of days, nearly identical spin-up and 
spin-down rates, and very small changes in X-ray luminosity. We propose that 
this phenomenon is the result of sudden dynamical changes in the accretion 
disks triggered by a gradual variation of mass accretion rates. 
These sudden torque reversals may 
occur at a critical accretion rate $\sim$ $10^{15}-10^{16}g~s^{-1}$ when the
system makes a transition from (to) a primarily Keplerian flow to (from)
a substantially sub-Keplerian, radial advective flow in the inner disk. 
For systems near spin equilibrium, the spin-up torques in the Keplerian
state are slightly larger than the spin-down torques in the advective state,
in agreement with observation. 
The abrupt reversals could be a signature of pulsar systems near spin 
equilibrium with the mass accretion rates modulated on a time scale 
of a year or longer near the critical accretion rate. It is interesting 
that cataclysmic variables and black hole soft X-ray transients change their
X-ray emission properties at accretion rates similar to the pulsars' critical
rate. We speculate that the dynamical change in pulsar systems shares a common 
physical origin with white dwarf and black hole accretion disk systems.

\end{abstract}


\section{Introduction}

Detailed spin evolution on long time scales has been made available 
for several accretion-powered X-ray pulsars such as 4U 1626-67, 
GX 1+4, and OAO 1657-415 (for comprehensive reviews and data, see e.g.
Chakrabarty et al. 1993, Chakrabarty 1995, and references therein). 
These systems have shown puzzling
abrupt spin reversals. Before and after the observed torque reversals,
their spin-up and spin-down torques were largely steady.
It is intriguing that the spin-up and spin-down torques before and after 
reversals are nearly identical. These reversals are quite different from 
the random torque fluctuations seen in some pulsar systems believed to be
fed by winds (e.g. Nagase 1989, Anzer \& B{\"o}rner 1995, and references
therein). The nearly steady torques plausibly indicate the existence of 
ordered accretion disks. The observations indicate that the mass accretion rate
is gradually modulated with small amplitudes on a time scale of at least
a year, which is much longer than the typical reversal time scale.

In the disk-magnetosphere interaction models of the Ghosh-Lamb type
(Ghosh \& Lamb 1979ab, Campbell 1992, Yi 1995, Wang 1995),
the magnetic torque is a function of the mass accretion rate, ${\dot M}$. 
The sign of the torque is reversed (spin-up/down) as ${\dot M}$ varies
when the disk inner edge moves past the equilibrium radius at which the 
torque vanishes (e.g. Lipunov 1992). In this picture, however, the torque 
variation is expected to be smooth and continuous unless ${\dot M}$ varies
discontinuously. Although this behavior may be relevant for some 
smooth torque reversals, the observed sudden reversals appear distinct
(Chakrabarty 1995). Given the lack of any plausible mechanism for 
discontinuous change of ${\dot M}$, which must be tuned to occur near 
spin equilibrium, it is difficult for the existing magnetized disk models 
to provide an explanation. The observations indicate that the mass accretion 
rates vary little during transition (cf. Chakrabarty 1995), 
which makes the discontinuous change of ${\dot M}$ unlikely as an explanation.

In this Letter, we propose a possible explanation for the observed sudden 
torque reversal. We take the neutron star moment of inertia 
$I_*=10^{45} g~cm^2$, radius $R_*=10^{6} cm$, and mass $M_*=1.4M_{\odot}$. 
At a cylindrical radius $R$ from the star, the vertical component of
the dipole magnetic field $B_z(R)=B_*(R_*/R)^3$ where $B_*$ is the stellar 
surface field strength. The spin period is $P_*$ or angular velocity 
$\Omega_*=2\pi/P_*$.

\section{Keplerian Disk-Magnetosphere Interaction and Inner Region}

In the conventional disk-magnetosphere interaction model of the Ghosh-Lamb
type (Ghosh \& Lamb 1979ab), the magnetic field of an accreting neutron
star penetrates a geometrically thin accretion disk and exerts a magnetic
torque. Except in a narrow region near the radius where the disk is disrupted,
it is assumed that the accretion disk rotation is Keplerian,
$\Omega_{K}(R)=(GM_*/R^3)^{1/2}$, the radial internal pressure 
gradient is small, the radial drift velocity is small, and the disk thickness 
becomes negligible (Campbell 1992, Yi 1995, Wang 1995 and references therein). 
In such a model, the $\phi$ component of the induction equation in steady 
state gives the azimuthal component of the field
\beq
B_{\phi}(R)={\gamma\over\alpha}{\Omega_*-\Omega_{K}(R)\over\Omega_{K}(R)}B_z(R).
\eeq
where we have assumed that the internal viscosity and the magnetic diffusivity 
are due to a single turbulent process.  Here we will assume
that $\gamma$, defined as the ratio of $R$ to the vertical velocity shear 
length scale $\left|v_{\phi}/(\partial v_{\phi}/\partial z)\right|$,
is $\sim 1$ (e.g. Campbell 1992, Yi 1995). 
The parameter $\alpha$ is the usual viscosity parameter (e.g. Frank et al.
1992) and we take $\alpha=0.3$. 
Although we adopt specific values of $\gamma$ and $\alpha$, the constraints on
$B_*$ could always be rescaled in such a way that the exact individual values 
of the three parameters are not necessary (e.g. Kenyon et al. 1996). 
The inner edge of the 
disk, where the disk is magnetically disrupted, is at $R=R_o$ determined by
the condition that the magnetic torque exceeds the internal torque in the
disk (Campbell 1992, Yi 1995, Wang 1995) which can be expressed as
\beq
\left(R_o\over R_c\right)^{7/2}={2N_c\over {\dot M}(GM_*R_c)^{1/2}}
\left[1-\left(R_o\over R_c\right)^{3/2}\right]
\eeq
where $N_c=(\gamma/\alpha)B_c^2R_c^3$, $B_c=B_z(R=R_c)$, and
$R_c=(GM_*P_*^2/4\pi^2)^{1/3}$ is the Keplerian corotation radius.
Integrating the magnetic torque over the disk, and allowing for the
angular momentum carried by the gas which crosses the inner edge of
the disk, the torque exerted on the star by the disk is
\beq
N={7\over 6}N_0{1-(8/7)(R_o/R_c)^{3/2}\over 1-(R_o/R_c)^{3/2}}
\eeq
where $N_0={\dot M}(GM_*R_o)^{1/2}$ (Campbell 1992, Yi 1995, Wang 1995) and
$N\rightarrow 0$ as $R_o/R_c$ approaches the equilibrium point $x_{eq}=0.915$. 
This estimate is based on the assumption that there is a negligible
flux of angular momentum from $R<R_o$.
Most of the contribution to the torque in this type of model comes from 
field-disk interaction in a narrow region just outside of $R_o$.
Several different phenomenological descriptions of disk-magnetosphere 
interaction cause little practical differences to our conclusions (Wang 1995).

Fig. 1(a) presents an example of the typical smooth torque reversal expected 
in the models of the Ghosh-Lamb type is shown. 
This example is based on a set of parameters 
similar to what we adopt below for 4U 1626-67. It is clear that any densely 
sampled spin evolution would reveal a gradual and continuous variation of the 
torque, a generic prediction of the Ghosh-Lamb type model. A sudden torque 
reversal with nearly constant $\left|N\right|$ is hard to explain 
unless there exists an 
unknown constraining mechanism or the ${\dot M}$ variation is discontinuous
and fine-tuned. It is possible that some observed smooth torque 
reversals (Chakrabarty 1995) could be due to this type of reversal model.

Under suitable conditions the inner parts of the accretion disk may
evolve to an optically thin, low density state,  for example,
inside the disk boundary layer around a white dwarf (e.g. Paczynski 1991, 
Narayan and Popham 1993).  When $\dot M$ is sufficiently small this
transition may occur over a broad range of radii within a disk, producing
a state in which radiative losses are so inefficient that the disk
retains a large fraction of the heat generated from the dissipation of 
orbital energy (the `advective state', cf. Narayan and Yi 1995).  
The transition to this state is not well understood,
and may be affected by various external factors, including 
X-ray irradiation (e.g. Meyer \& Meyer-Hofmeister 1990) and coronal
heating (Meyer \& Meyer-Hofmeister 1994).

Here we will assume that, whatever the details of the transition, the
disk will make the jump to a low density state when it becomes possible
for it to do so. What is the critical ${\dot M}_{crit}$ below which the disk 
becomes hot and optically thin? Observationally, weakly magnetized cataclysmic 
variables generally show a trend in which the X-ray to optical 
flux ratio decreases as ${\dot M}$ increases. Above a critical rate of 
$\sim 10^{16} g~s^{-1}$,
the X-ray emission becomes extremely weak. This has been attributed to the
transition of the inner region to an optically thin (X-ray emitting) hot
accretion disk (Patterson \& Raymond 1985, Narayan \& Popham 1993) at low
${\dot M}$. Interestingly, such a critical rate is largely consistent
with the critical rate $\sim 2\times 10^{17}\alpha^2$ or $\sim 2\times 10^{16}
g~s^{-1}$ for $\alpha\sim 0.3$ based on the recently discussed 
advection-dominated hot accretion disks (Narayan \& Yi 1995). 
For a given accretion rate, there exists a critical radius inside of which 
the disk makes a transition to sub-Keplerian while the disk at radii larger than
the critical radius remains Keplerian (Narayan \& Yi 1995). 
The exact location of this critical radius is not clearly understood yet.
We assume that most of the magnetic torque contribution comes from the
inner region (e.g. Wang 1995) which is inside the critical radius,
e.g. $R_{crit}\gg R_c>R_o$.
The relevance of the critical ${\dot M}_{crit}$ for the cataclysmic
variables becomes more striking when we consider strongly magnetized neutron
star systems. The inner edge of the accretion disk around a strongly 
magnetized neutron star lies roughly at
$R_o\sim 5\times 10^{8}\left(B_*/10^{12}G\right)^{4/7}
\left({\dot M}/10^{16}g/s\right)^{-2/7} cm$
which is close to the typical white dwarf radius. Therefore, one may ask what 
would be the effects of the transition,
at ${\dot M}_{crit}\sim 10^{16}g~s^{-1}$, 
to the hot optically thin accretion disk on the spin evolution of the pulsars?

\section{Disk Transition and Torque Reversal}

We propose an explanation for the torque reversal based on such a transition. 
The optically thin, hot accretion disk cannot be geometrically thin or 
Keplerian once its internal pressure $\sim\rho c_s^2$ becomes a significant
fraction of its orbital energy. 
After the transition, the disk thickness $H\sim c_s/\Omega_K$ and radial 
drift velocity $v_R\sim\alpha c_s^2/\Omega_K$ increase. The rotation of the 
accretion disk $\Omega$ becomes sub-Keplerian $\Omega<\Omega_K$ 
(e.g. Narayan \& Yi 1995). In the case of strongly magnetized pulsars, 
direct X-ray observation of such a disk transition is 
difficult because most of the X-ray luminosity, $L_x\sim GM_*{\dot M}/R_*$,
comes from the surface of the star and the luminosity from the disk, which
is truncated well above the stellar surface, is limited to
$\sim GM_*{\dot M}/R_o\ll L_x$. The sub-Keplerian disk, however, may have 
observable dynamical consequences. Once the sub-Keplerian rotation is 
forced on the magnetosphere, the (sub-Keplerian) corotation radius is shifted 
inward and the position of the disk inner edge with respect to the new 
corotation radius is relocated. (In principle, the inner edge of the disk
could end up beyond the new corotation radius but this would not account for
the observed properties of the torque reversal systems.)
As a result, the magnetic torque changes and there could be a visible change 
of spin-up/down torque on the disk transition time scale.

If such a transition does occur at a certain critical rate, 
the most likely time scales are the local thermal time scale 
$t_{th}\sim (\alpha\Omega_K)^{-1}$ or the disk viscous-thermal time scale 
$t_{dis}\sim R/(\alpha c_s)\sim (R/H)(\alpha\Omega_K)^{-1}\sim 10^3s$
for $\alpha\sim 0.3$, $R\sim 10^9cm$, and ${\dot M}\sim 10^{16} g~s^{-1}$
(e.g. Frank et al. 1992). This is plausible if the transition is mainly 
driven by the thermal instability of the local optically thin region of 
the accretion flow accompanied by the disk density change on the local 
viscous-thermal time scale (cf. Meyer \& Meyer-Hofmeister 1994). 
In fact, the viscous time scale for the thin disk is unrealistically long, 
since the relevant infall time is the one for the hot, and thick disk, 
which will be only a few orbital times. In any case, these time scales 
indicate that the local disk surface density change could occur on a time 
scale short enough, much less than a day, to make the transition appear 
almost instantaneous. The long term gradual ${\dot M}$ 
modulation determines the overall evolutionary trend and possibly affects the 
residuals seen in some observations (Cutler et al. 1986, Chakrabarty 1995). 
The sudden torque reversal does not require any short term 
($\sim day$) change of ${\dot M}$ but only that the condition
${\dot M}\sim {\dot M}_{crit}$ be satisfied around the time of reversal.

In order to model the reversal episode, we assume no dynamic vertical motion 
of the disk gas such as winds or outflows. 
The dominant effect of the transition is to reset 
the corotation radius and disk truncation radius $R_o$.
We take the temperature of the sub-Keplerian hot disk to be a constant fraction 
$\xi$ of the local virial temperature, i.e. $c_s^2\sim\xi R^2\Omega_K^2$
(e.g. Narayan \& Yi 1995). Then the ratio of the disk thickness to 
radius is $H/R\sim\xi^{1/2}$
and the radial drift velocity is $v_R\sim -\alpha\xi R\Omega_K$. 
Using the radial 
component of the momentum equation (e.g. Campbell 1992), 
we get the sub-Keplerian rotation frequency
$\Omega/\Omega_K
\sim\left(1-5\xi/2-\alpha^2\xi^2/2\right)^{1/2}\equiv A.
$
where $\xi\rightarrow 0$ corresponds to the usual Keplerian limit. 
We note that $v_R/R\Omega_K\sim \alpha\xi <1$ and $H/R\sim \sqrt{\xi}\le 1$. 
Assuming a constant $\xi$ or $A$, after the transition to the sub-Keplerian 
rotation with $\Omega(R)=A\Omega_K(R)<\Omega_K(R)$, 
the corotation radius becomes $R_c^{\prime}=A^{2/3}R_c$ and the new inner 
disk edge is relocated to $R_o^{\prime}$ determined by
\beq
\left(R_o^{\prime}\over R_c^{\prime}\right)^{3}={2N_c\over N_0^{\prime} A}
\left[1-\left(R_o^{\prime}\over R_c^{\prime}\right)^{3/2}\right]
\eeq
where $N_0^{\prime}=A{\dot M}(GM_*R_o^{\prime})^{1/2}$.
The torque on the star after the transition is
\beq
{N^{\prime}\over N_0^{\prime}}={7\over 6}
{1-(8/7)(R_o^{\prime}/R_c^{\prime})^{3/2}\over
1-(R_o^{\prime}/R_c^{\prime})^{3/2}}.
\eeq
The torque vanishes when $R_o^{\prime}/R_c^{\prime}\rightarrow x_{eq}=0.915$
as in the Keplerian rotation (eq. 2-3). In our discussions, we take
a constant $A=0.2$ (e.g. Narayan \& Yi 1995). The parameters $\alpha=0.3$
and $\gamma=1$ are assumed to be constant before and after the transition.
For the Keplerian disk, the equilibrium spin ($N=0$ in eq. (2-3)) period is
\beq
P_{eq}=[4.9s]\left(\gamma\over\alpha\right)^{3/7}
\left(B_*\over 10^{12}G\right)^{6/7}\left(R_*\over 10^6cm\right)^{18/7}
\left(M_*\over 1.4M_{\odot}\right)^{-5/7}
\left({\dot M}\over 10^{16}g/s\right)^{-3/7}.
\eeq
For $\Omega=A\Omega_K<\Omega_K$, the equilibrium spin period would become 
longer by a factor $1/A$ and the system begins to evolve toward the newly 
determined equilibrium after transition. 

\section{Sudden Torque Reversals}

For the spin evolution calculation we integrate the torque equation
\beq
{dP_*\over dt}=-{P_*^2\over I_*}[{\rm Torque}]
\eeq
where Torque $=N$ or $N^{\prime}$ depending on the physical state of the inner
disk. We take a linear increase or decrease of ${\dot M}$ as an approximation 
to more complex ${\dot M}$ variations on longer time scales. 
The transition occurs on a time scale 
$\ll P_*/\left|dP_*/dt\right|$ before and after the transition. 
The transition at ${\dot M}_{crit}=10^{15}-10^{16} g/s$, which is determined 
by the fits to the observed spin evolution, is taken to be instantaneous
(cf. $t_{th}$, $t_{dis}$).
For each torque reversal event, we adjust $B_*$, ${\dot M}$, and the accretion 
rate time scale, ${\dot M}/\left|d{\dot M}/dt\right|$.
We consider three X-ray pulsars for which abrupt torque reversals have been 
detected. For a given initial spin period $P_*$, a fit gives a set of the 
above parameters.

{\it 4U 1626-67:} 
4U1626-67 ($P_*\approx 7.7s$) was steadily spun-up on a time scale 
$\sim 10^4 yr$ (${\dot P_*}/P_*^2=-8.54(7)\times 10^{-13} s^{-2}$)
during 1979-1989. The Keplerian corotation radius $R_c\approx 
(GM_*P_*^2/4\pi^2)^{1/3}=6.5\times 10^8 cm$. The recent
BATSE detection of a sudden torque reversal to spin-down is puzzling due to
its very short time scale and the nearly equal spin-up/down rates.
The steady spin-down torque suggests that there remains a dynamically stable 
(disk) structure after the sudden reversal (For details of observations, see
Chakrabarty 1995).
In Fig. 1(b), the observed torque reversal event is reproduced by 
$B_*=1.2\times 10^{11}G$, ${\dot M}=4\times 10^{15} g/s$, and 
$d{\dot M}/dt=-5\times 10^{13} g/s/yr$ which give $R_o/R_c=0.58$,
$R_o^{\prime}/R_c^{\prime}=0.95$, and $R_o^{\prime}/R_o=0.56$. 
The derived accretion rate
is slightly lower than the previously quoted values (Chakrabarty 1995),
and accordingly the derived $B_*$ is also lower than the previous estimates 
(e.g. Pravado et al. 1979, Kii et al. 1986, Chakrabarty 1995). We note, however,
that our estimated values ($B_*$ and ${\dot M}$) can always be rescaled by 
changing $\alpha$ and $\gamma$ (Kenyon et al. 1996). The values of $\alpha$
and $\gamma$ based on first principles are not available.
The gradual decrease of ${\dot M}$ is consistent with the observed flux
decrease (Mavromatakis 1994). The fit naturally achieves the spin-down torque
which is slightly smaller than the spin-up torque. 
The fit requires for the 
transition to occur at ${\dot M}_{crit}=3.3\times 10^{15} g~s^{-1}$. 
The gradual decrease of the mass accretion rate on a time scale $\sim
20 yrs$ cannot be due to any viscous or thermal processes
operating in the inner region ($t_{th}$, $t_{dis}$). 

{\it OAO 1657-415:}
OAO 1657-415 has an observed pulse period $P_*\approx 38s$. Recent observed
spin-up/down torques are ${\dot P}_*/P_*^2\approx -7\times 10^{-12} s^{-2}$ and 
$\approx 2\times 10^{-12} s^{-2}$ respectively 
(Chakrabarty et al. 1993). For the torque reversal in Fig. 1(c), we get
$B_*=10^{12}G$, ${\dot M}=2.0\times 10^{16} g/s$, and
$d{\dot M}/dt=-5\times 10^{16} g/s/yr$. The characteristic ${\dot M}$
modulation time scale is $\sim 0.3yr$.
The critical accretion rate ${\dot M}_{crit}=1.1\times 10^{16} g/s$
which is somewhat higher than the value required for 4U 1626-67.

{\it GX 1+4:}
GX 1+4 has recently shown a sudden transition from spin-down to spin-up
around the spin period $P_*\approx 122s$ (Chakrabarty 1995 and references
therein). GX1+4 is peculiar in the sense that despite its very short spin 
time scale $\sim 40 yr$, the spin equilibrium has not been reached.
It is likely that ${\dot M}$ fluctuates or oscillates on a time scale 
$\ll 40yr$ near spin equilibrium.  The spin-down rate in the 1980's, 
${\dot P_*}/P_*^2\sim 3.7\times 10^{-12} s^{-2}$, is not far from the 1970's 
spin-up rate, ${\dot P_*}/P_*^2\sim -6.0\times 10^{-12} s^{-2}$. The recent 1994
torque reversal from spin-down to spin-up lasted for $\sim 100d$.
This system also showed very similar spin-down and spin-up rates.
Although there is no significant spectral change in the hard X-ray emission
spectra during the spin evolution, the flux appears to be increasing as 
spin-down torque increases (Chakrabarty 1995), 
which is in contradiction to the behavior
expected in the Ghosh-Lamb type model (eqs. (2-2),(2-3)).
The fit shown in Fig. 1(d) corresponds to
$B_*=3.2\times 10^{12} G$, ${\dot M}=5.0\times 10^{15} g/s$,
and $d{\dot M}/dt=1.0\times 10^{16} g/s/yr$. Once again, we note that
these parameters could be rescaled by different choices of $\alpha$ and 
$\gamma$. It is interesting to observe 
that the ${\dot M}$ modulation time scale ${\dot M}/({\dot M}/dt)\sim 5yr$ is 
not far from the detected time scale in the coherent variation of pulse 
frequency residual (Cutler et al. 1986, Chakrabarty 1995). The critical mass 
accretion rate ${\dot M}_{crit}=6.5\times 10^{15} g/s$ lies between the
two values derived above. The gradual decrease of the spin-up torque after 
reversal is not accurately fit in Fig. 1(d) 
with the linear ${\dot M}$ variation. 
This is not surprising given the reported unsteady behavior of the 
spin-down torque before the reversal (Chakrabarty 1995). 
The observed $\sim 300d$ coherent variation in pulse frequency residual is
intriguing but such a time scale is not far from the estimated orbital time 
scale $\sim yr$ (Chakrabarty 1995).

\section{Summary and Discussions}

The proposed transition is most likely to occur at a critical accretion rate 
${\dot M}_{crit}\sim 10^{15}-10^{16}g/s$.
This suggests an interesting connection between the pulsar systems
and other compact accretion systems such as cataclysmic variables and 
black hole soft X-ray transients. We speculate that the transitions seen
in these systems may be due to a common physical mechanism, i.e. disk 
transition to optically thin hot flow. The model indicates that the sudden 
torque reversal could be a signature of a pulsar system near spin-equilibrium 
with ${\dot M}\sim {\dot M}_{crit}$.

There are some outstanding issues to look into.
(i) The origin of the gradual ${\dot M}$ modulation on a time scale ranging
from $\sim yr$ to a few decades remains unknown. The modulation by
the orbital motion on a time scale $\sim yr$ is plausible in GX 1+4
(Chakrabarty 1995) but is unlikely in 4U 1626-67 (Rappaport et al. 1977, 
Joss et al. 1978, Shinoda et al. 1990, Chakrabarty 1995) and 
OAO 1657-415 (Chakrabarty et al. 1993). Several binary precession time scales 
(e.g. Thorne et al. 1986) and the mass flow oscillation time scale 
(induced by X-ray irradiation, Meyer \& Meyer-Hofmeister 1990) could be
relevant for long time scale ($>month$) modulations. 
In 4U 1626-67 the observed optical and X-ray pulsation frequencies are
identical, which has been attributed to the reprocessing of X-rays 
by accretion disk (Ilovaisky et al. 1978, Chester 1979).
The direct X-ray irradiation of the secondary star
(Hameury et al. 1986) and the disk instability (e.g. Smak 1984) may also
give rise to ${\dot M}$ modulations in neutron star systems.
(ii) It is important to quantitatively understand ${\dot M}_{crit}$ and 
$R_{crit}$ (cf. Narayan \& Yi 1995). 
(iii) Within our model, the observed smooth transitions back to spin-up (seen
in OAO 1657-415 and GX 1+4) could result from the return from the 
advective to Keplerian flow. 
The characteristic time scale for such a back transition is
likely to be the the viscous disk formation time scale. The observed UV-delay 
time scale, on the order of a day, in cataclysmic variables (Livio \& Pringle 
1992), may be similar to the postulated reverse transition time scale. 
(iv) It remains unexplained that in GX 1+4 the X-ray flux increased during 
the increase of the spin-down torque (Chakrabarty 1995). 
If X-rays come from the shocked polar accretion, 
as ${\dot M}$ decreases, the X-ray emission temperature and the apparent
flux in a fixed X-ray band could decrease due to radiation drag
(e.g. Yi \& Vishniac 1994). A geometrically thick, but 
optically thin, inner disk is more apt to scatter X-ray emission, 
so that a partially obscured polar cap can actually become more conspicuous 
as $\dot M$ drops.

\acknowledgments
This research was supported in part by NSF Grant 95-28110 (JCW), by the
SUAM Foundation (IY), and by NASA grant NAG5-2773 (ETV). We
are happy to acknowledge related
discussions on pulsars with Josh Grindlay and Ramesh Narayan.

\vfill\eject

\centerline{Figure Caption}

\noindent
Figure 1:
(a) A typical smooth torque reversal event expected in the Ghosh-Lamb type 
model. This illustrative example assumes $B_*=10^{12} G$, 
$P_*=7.68s$ and ${\dot M}$ linearly decreasing from $7.2\times 10^{16} g/s$ to 
$5.0\times 10^{16} g/s$ over 10 yrs. The torque passes through $N=0$ 
(eq. (2-3)) and the continuous variation should show a wide range of torques 
before and after the torque reversal. Upper panel: time variation of 
torque. Lower panel: spin-up to spin-down transition.
Examples of (b) 4U 1626-67, (c) OAO 1657-415, and (d) GX 1+4. 
The solid lines correspond to
models described in the text and the dashed lines connect data points 
adopted from Chakrabarty et al. (1993) and Chakrabarty (1995). In OAO 1657-415
and 4U 1626-67, the observed spin periods are shown only schematically.

\vfill\eject\clearpage

\centerline{\bf ERRATUM}

In the Letter "Torque Reversal in Accretion-Powered X-ray Pulsars" by I. Yi,
J. C. Wheeler, \& E. T. Vishniac (ApJ, 481, L51 [1997]), 
there are errors in the values of the
magnetic fields and the mass accretion rates. The correct values are as 
follows and Figure 1 is revised. {\bf 4U 1626-67:} $B_*=6\times 10^{11}G$, 
${\dot M}=2.8\times 10^{16} g/s$, $d{\dot M}/dt=-4.5\times 10^{14} g/s/yr$, 
$R_o/R_c=0.760$, $R_o^{\prime}/R_c^{\prime}=0.987$, $R_o^{\prime}/R_o=0.444$,
${\dot M}_{crit}=2.2\times 10^{16} g/s$. 
{\bf OAO 1657-415:} $B_*=5\times 10^{12}G$, ${\dot M}=1.2\times 10^{17} g/s$, 
$d{\dot M}/dt=-1.2\times 10^{17} g/s/yr$, 
${\dot M}_{crit}=1.0\times 10^{17}g/s$. 
{\bf GX 1+4:} $B_*=1.3\times 10^{13}G$, ${\dot M}=4\times 10^{16} g/s$,
${d{\dot M}/dt}=1.3\times 10^{17}g/s/yr$, ${\dot M}_{crit}=6\times 10^{16} g/s$.

\vskip0.5cm

\centerline{Figure 1-Corrected}
Torque reversal events in three X-ray pulsar systems. 
Short dashed lines correspond to the smooth torque transition (see text)
and thick solid lines correspond to the sudden torque transition proposed
in the present work. The points connected by long dashed lines are
observed data points adopted from Chakrabarty et al. (1993) and Chakrabarty
(1995). In OAO 1657-415 and 4U 1626-67, the observed spin periods are shown
only schematically. Examples of (a) 4U 1626-67, (b) OAO 1657-415, and 
(c) GX 1+4. For the smooth transition events, the parameters ($B_*, {\dot M},
d{\dot M}/dt$) are ($10^{12}G$,$~4.2\times 10^{16}g/s$,
$~-2\times 10^{15}g/s/yr$) for 4U 1626-67, ($10^{13}G$,$~2\times 10^{17}g/s$,
$~-5\times10^{17}g/s/yr$) for OAO 1657-415, and ($4\times 10^{13}G$,
$~10^{15}g/s$, $~1.8\times 10^{17}g/s/yr$) for GX 1+4.

\vfill\eject


\begin{references}

\reference Anzer, U. \& B{\"o}rner, G. 1995, A\&A, 299, 62 
\reference Campbell, C. G. 1992, Geophys. Astrophys. Fluid Dyn., 63, 179
\reference Chakrabarty, D. 1995, Ph.D. Thesis, California Institute of 
Technology
\reference Chakrabarty, D. et al. 1993, ApJ, 403, L33
\reference Chester, T. J. 1979, ApJ, 227, 569
\reference Cutler, E. P., Dennis, B. R., \& Dolan, J. F. 1986, ApJ, 300, 551
\reference Frank, J., King, A. R., \& Raine, D. 1992, Accretion Power in 
Astrophysics (Cambridge: Cambridge University Press)
\reference Ghosh, P. \& Lamb, F. K. 1979a, ApJ, 232, 259
\reference Ghosh, P. \& Lamb, F. K. 1979b, 234, 296
\reference Hameury, J.-M., King, A. R., \& Lasota, J.-P. 1986, A\&A, 162, 71
\reference Ilovaisky, S. A., Motch, C., \& Chevalier, C. 1978, A\&A, 70, L19
\reference Joss, P. C., Avni, Y., \& Rappaport, S. 1978, ApJ, 221, 645
\reference Kenyon, S. J., Yi, I., \& Hartmann, L. 1996, ApJ, 462, 439
\reference Kii, T., Hayakawa, S., Nagase, F., Ikegami, T., \& Kawai, N. 1986,
PASJ, 38, 751
\reference Lipunov, V. M. 1992, Astrophysics of Neutron Stars 
(Berlin: Springer-Verlag)
\reference Livio, M. \& Pringle, J. E. 1992, MNRAS, 259, 23
\reference Mavromatakis, F. 1994, A\&A, 285, 503
\reference Meyer, F. \& Meyer-Hofmeister, E. 1990, A\&A, 239, 214
\reference Meyer, F. \& Meyer-Hofmeister, E. 1994, A\&A, 288, 175
\reference Nagase, F. 1989, PASJ, 41, 1
\reference Narayan, R. \& Popham, R. 1993, Nature, 362, 820
\reference Narayan, R. \& Yi, I. 1995, ApJ, 452, 710
\reference Paczynski, B. 1991, 370, 597
\reference Patterson, J. \& Raymond, J. C. 1985, ApJ, 292, 535
\reference Pravado, S. H. et al. 1979, ApJ, 231, 912
\reference Rappaport, S. et al. 1977, ApJ, 217, L29
\reference Shinoda, K. et al. 1990, PASJ, 42, L27
\reference Smak, J. 1984, PASP, 96, 5
\reference Thorne, K. S., Price, R. H., \& MacDonald, D. A. 1986, Black Holes:
The Membrane Paradigm (New Haven: Yale University Press)
\reference Wang, Y.-M. 1995, ApJ, 449, L153
\reference Yi, I. 1995, ApJ, 442, 768
\reference Yi, I. \& Vishniac, E. T. 1994, ApJ, 435, 829

\end{references}
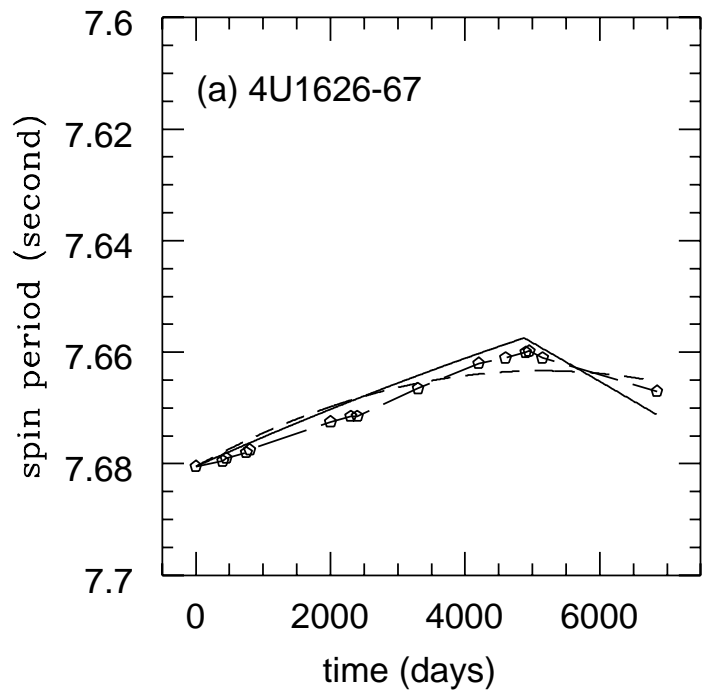
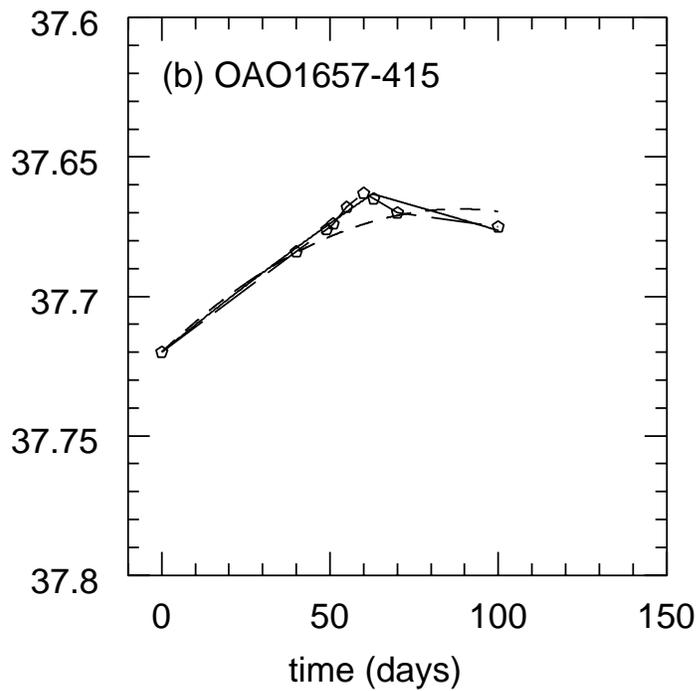
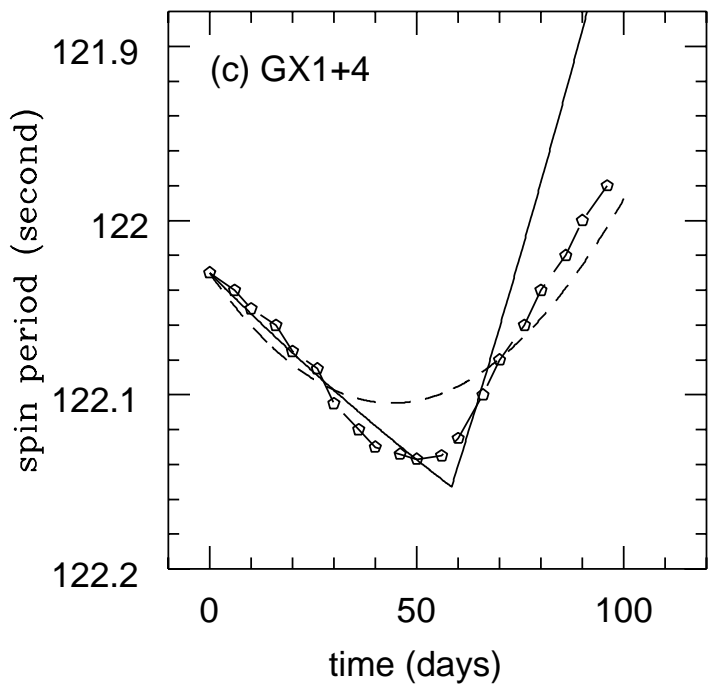
\end{document}